\documentstyle[aps,citesort,preprint,epsfig]{revtex}
 \tightenlines

\newcommand{\angstrom}{{\rm \AA}}
\newcommand{\atm}{{\rm \,atm}}
\newcommand{\kelvin}{{\rm \,K}}

\begin{document}

\title{
        Computer Simulations of the Wetting Properties \\ 
 	of Neon on Heterogeneous Surfaces}
\author{Stefano Curtarolo$^{1,3,4}$, Milton W. Cole$^1$, Mary J. Bojan$^2$, and William A. Steele$^2$}
\address{ 
$^1$Department of Physics, Penn State University, University Park, PA 16802, USA \\
$^2$Department of Chemistry, Penn State University, University Park, PA 16802, USA \\
$^3$University of Padua, Physics Department, Padua, Italy \\   
$^4${\em e-mail address: auro@pd.infn.it}\\ 
{\em phone: (814)863-8460, fax: (814)863-5319} }

\date{\today}
\maketitle
\begin{abstract}
{\sf We use the grand canonical Monte Carlo method to study the nature of
wetting transitions on a variety of heterogeneous surfaces.  The model
system we explore, Ne adsorption on Mg, is one for which a prewetting
transition  has been found in our previous simulations. We find that the
first order transition present on the flat surface is absent from the rough
surface. Nevertheless, the resulting isotherms are, in some cases, so close
to being discontinuous that the distinction would be difficult to discern
in most experiments.}

\end{abstract}
\newpage


\section{Introduction}

The last decade has seen a great increase in the study of wetting
transitions. Particular attention has been drawn to the case of simple
gases on alkali metals surfaces, for which the first experimental examples
of the prewetting (first order transition) phenomenon have been seen
\cite{taborek_rutledge_1992,rutledge_taborek_1992,ketola_wang_hallock_1992,cheng_mistura_lee_chan_cole_carraro_saam_toigo_1993,taborek_rutledge_1993,mistura_lee_chan_1994,wyatt_klier_stefanyi_1995,hallock_1995}. 
The subject has attracted even wider interest with the recent addition of Hg
transitions to the list of  adsorption systems exhibiting these phenomena
\cite{kozhevnikov_arnold_naurzakov_fisher_prl_1997,hensel_yoa_ejssic_1997}.
Theoretical interest in this problem is extensive and diverse; important
open questions include which systems are likely to display the transitions
\cite{ebner_and_saam_1977,cahn_1977_2,soko_fisher_1990,fan_monson_1993,nijmeijer_bruin_bakker_van_leeuven_1991},
their sensitivity to the adsorption potential
\cite{bojan_cole_johnson_steele_wang_1998,cole_fpe_1998},
the nature and dimensionality of long-range force \cite{lipowsky_prb_1985},
the possibility of higher order transitions \cite{tarazona_evans_ml_1983}, 
and the influence on the transition of surface
irregularity 
\cite{hui_berker_1989,aizenman_wehr_1989,berker_jap_1991,berker_pa_1993,bruin_pa_1998,blossey_kinoshita_muller_dupontroc_jltp_1998}.

In a previous paper \cite{bojan_cole_steele_to_be_published}, 
henceforth called I, we explored several of these
questions.  Specifically, we used the grand canonical Monte Carlo ({\sf gcmc})
method of statistical mechanics to compute the nature of wetting
transitions of Ne on weakly adsorbing surfaces. That work found that alkali
metals attract Ne so weakly that nonwetting behavior was predicted for all
temperatures below $42\,\kelvin$, i.e. $95 \%$ of the Ne bulk critical temperature.
This result is consistent with experimental data of Hess, Sabatini, and
Chan; using a quartz microbalance technique, they found a wetting
transition near $43.4\,\kelvin$ on Rb and evidence of a drying transition on a Cs
surface \cite{hess_sabatini_chan_1997}.

In I, we found that the case of Ne adsorption on Mg is quite different from
that on the alkali metals; the reason is that the Ne adsorption potential
on Mg is approximately four times as attractive as that experienced  on Cs
and twice as attractive as that on Li \cite{chizmeshya_and_cole_and_zaremba_1998}. 
We found prewetting transition behavior on Mg in the regime $22 < T < 30\,\kelvin$. 
This is manifested as a coverage
discontinuity at a pressure which depends sensitively on both T and the
adsorption potential.  In paper I, we also explored the effect on the
transition of a (periodic) corrugation of the adsorption potential  by
constructing a simple cubic model of Mg, with properties chosen to
correspond qualitatively to real Mg (e.g. lattice constant $a=4.01\,\angstrom$).
We found that the principal effect of this corrugation was a small shift in
the prewetting transition characteristics, as expected from a qualitative
argument which attributes the shift to the atoms' extra attraction due to
the periodic part of the potential.

This paper extends the Ne/Mg study to the case of heterogeneous surfaces.
Such surfaces are here constructed by either adding Mg atoms to or
subtracting  them from the semi-infinite simple cubic lattice employed in
I. We find a qualitative change in the adsorption behavior: the prewetting
transition disappears, as expected from general theory of adsorption on
irregular surfaces. In the case of small heterogeneity, we find that the
adsorption can exhibit a very rapid rise as a function of $P$; this
``quasi-transition'' might not be distinguishable from a true (first order)
transition in a laboratory experiment. In other cases, the rapid rise is
replaced by a fairly gentle increase of coverage with $P$. As $P$ approaches
saturated vapor pressure (svp), the coverage dependence may correspond to
either complete wetting (divergent film growth) or nonwetting behavior,
depending on the kind of heterogeneity. Thus we find that surface
irregularity can change a wetting system into a nonwetting system. This can
be rationalized from a crude thermodynamic argument based on the
interfacial free energy cost of an irregular film.

This paper presents our geometry and computational method in Section 2 and
our results in Section 3. We discuss our  conclusions and open questions in
Section 4.  We emphasize that our results may well depend on the specific
model, which is not completely general.

\section{Geometry and method}

Our calculational technique is discussed extensively in I, to which we
refer the interested reader. Briefly, we perform gcmc simulations of Ne
adsorption on a model surface. The Ne atoms are confined to a space bounded
on one side by the surface and on the other by a region ($z > L= 78\,\angstrom$)
of infinite potential energy.  The Mg surface is periodically replicated
as discussed below. It gives a semi-infinite domain of Mg
atoms which reside on sites of a simple cubic lattice. Each Mg atom
interacts with the Ne atoms with a Lennard-Jones (LJ) pair potential with
parameters $\epsilon_{gs}=15\,\kelvin$ and $\sigma_{gs}=5.01\,\angstrom$. 
These  values were chosen in I to approximate the theoretical potential of 
Chizmeshya {\it et al} \cite{chizmeshya_and_cole_and_zaremba_1998}, appropriate to
the case of a Ne atom above a flat Mg surface. Finally, the Ne-Ne
interaction is also taken to have  LJ  form, with parameters $33.9\,\kelvin$ and
$2.78\,\angstrom$. These assumed functional forms, as well as the assumption of
classical statistical mechanics, should be of at least qualitative accuracy
and are conventional in this field. We do not aspire to quantitative
accuracy in the potential, because  (to our knowledge) no one really knows the
characteristics of the physisorption potential in the case of a rough
metallic surface \cite{foot_1}. 

The initially flat surface consists of ($x/y$ periodically replicated) square
cells of  dimension $28.07\,\angstrom$, each containing 49 surface atoms. The
algorithm used for creating our rough surfaces involves a surface profile
function produced in momentum space (and then Fourier transformed).  The
procedure begins with a random, Gaussian-distributed  profile function.
Then one applies a correlation filter to remove high wave vector
components. The filter  is a Lorentzian function, with a width equal to the
periodicity. Our rough surfaces are characterized in either of two ways.
One involves a quantity called $\sigma$, defined as the root mean square
deviation (from a mean value of zero) of the topmost atoms' values of $z$.
The other is the  specification of the rough geometry, i.e. the actual
number and shapes of the various imperfections.
We note that our ``rough'' surface is actually periodic, but this should not
affect our conclusions excepting those phenomena (i.e. exactly at transition) 
which involve fluctuations with wave lengths larger than the cell size 
(i.e. 28.07 \angstrom). The same limitation applies, of course, to all phenomena 
studied with simulations.

\section{Results}

Our methods and results involve fully three dimensional (3d) functions,
such as the density and potential energy. For ease of depiction and
interpretation, we present 2d graphs of different kinds. We define, in
general, a 2d function $U(x,y)$ as the minimum, as a function of $z$, of the 3d
potential $V(x,y,z)$ experienced by a Ne atom. An example appears in Fig. 1a,
which shows this function in the case of a simulation unit cell which
possesses both a single added Mg atom and a single atomic ``pit'', i.e. a hole
created by the extraction of one surface layer Mg atom. Fig. 1b shows the
dependence on $z$ of $V(x,y,z)$ above three different Mg atoms  on this surface
(an ordinary surface atom, an adatom, and an atom at the bottom of the
pit). One is struck by the fact that the Mg adatom creates a very extended
region of unfavorable potential for a Ne atom. This is a consequence of the
large hard core length of the Ne-Mg pair potential. In contrast, one
observes that the pit provides a very attractive  region; however, this
attractive region is very narrow and does not extend significantly above
the neighboring surface atoms. We shall see, as a  result, that the pit
does not greatly enhance Ne adsorption in its vicinity. Note also in Fig.
1a that away from the imperfections the function $U(x,y)$ varies by a factor
of about $20 \%$, due to the atomic periodicity alone. This variation and its
effect on wetting were discussed in I.

All of the simulations in this paper have been carried out at a temperature
$T = 28\,\kelvin$. This temperature is intermediate between the wetting temperature
$(22\,\kelvin)$ and the prewetting critical temperature $(30.6\,\kelvin)$ computed for the
(nonperiodic, flat)  Mg surface in I \cite{foot_2}.
Figure 2 displays the
isotherms computed for Ne  on several different Mg surfaces. For reference,
we note that a nominal monolayer coverage is about 60 Ne atoms per periodic
surface cell, derived by assuming a 2d density equal to the $2/3$ power of
the density of the 3d liquid. One of the isotherms in Fig. 2 is that
obtained in paper I with the ``perfect'' flat surface of Mg; one observes
there a vertical prewetting transition at $P= 0.862\,\atm$. This corresponds to
a coverage jump by a factor of four, as discussed in I. The closest
isotherm rise to this arises in the case of a small pit (created by
removing one surface atom). This appears to be discontinuous near
$P=0.93\,\atm$. We have not yet employed
sufficient computational resources to examine this functional dependence in
detail; the difficulty lies in the divergent fluctuations in coverage when
isotherms become so very steep. A nonanalytic dependence of $N$ on $P$ cannot
be ruled out at this time. A third curve in figure 2 corresponds to the
case of a single adatom. Here, too, we find a rapid variation of coverage
with P, occurring at somewhat higher $P=0.92\,\atm.$ Finally, one observes
another isotherm, corresponding to  both an adatom and a pit, i.e. the
geometry corresponding to Fig. 1. The rise here is even more rapid than in
the single adatom case, which is perhaps curious because the surface is
less homogeneous. Similarly surprising, at first glance, is that the rise
occurs at even higher $P=0.952\,\atm.$ than for the other two heterogeneous
cases. Very naively, one might have expected the isotherm for the case of
an adatom and a pit to be intermediate between those in the cases of a
single impurity, i.e. adatom or pit.  That expectation is a consequence of
a ``superposition'' supposition, which would be valid if the regions of
imperfection were remote from each other, so their effects did not
interfere. The rather different reality reflects the collective behavior of
the wetting transition; the net adsorption is not a superposition of
adsorption from distinct regions of the surface. Specifically, the case of
both an adatom and a pit is the least attractive overall of the four cases
considered, so that the adsorption rise occurs at the highest pressure of
all. When this quasi-transition ultimately occurs, the thick film (present
above the transition) experiences the remote heterogeneity weakly, so that
for $P$ higher than the quasi-transition value, the net coverage is almost
independent of the heterogeneity. This argument rationalizes the increased
abruptness of the adatom plus pit isotherm relative to that for the adatom
alone.

We define a 2d density in the usual way, by integrating the 3d density of
the film over the z coordinate at fixed $(x,y)$. Figs. 3 to 5 display the
evolution of this 2d density  at the quasi-transition in the case of a
simulation unit cell which possesses both an adatom and a  pit. One sees
that the adatom inhibits Ne adsorption near it almost entirely, even when
the (average) net coverage is several layers.  Note that the region at
lateral distance $6\,\angstrom$  from the adatom has instead a slightly enhanced
density due to the attractive potential (barely visible in Fig. 1a) at the
intersection of   the adatom and the rest of the surface. Finally the pit
very  slightly increases the adsorption in its vicinity (at all coverages).
This can be understood in terms of the attractive potential near the pit in
Fig. 1. Because this region of added attraction is so small, the net
contribution is only  a few per cent of monolayer coverage, i.e. a few Ne
atoms per unit cell of the simulation.

Fig. 6 shows the evolution of the adsorption behavior as the surface
becomes progressively rougher. This general behavior is straightforward to
interpret. Increasing irregularity forces the quasi-transition value of $P$
higher because the film growth is depressed by the irregularity. Note that
the quasi-transition in the cases of $\sigma$ equal to $0.2$ and $0.3\,\angstrom$ 
is nearly discontinuous. The reason is that the thick film, above the
transition, is insensitive to the heterogeneity below. Hence the dominant
effect of the latter is to postpone the jump to ever higher $P$. Ultimately,
at $\sigma=0.4\,\angstrom$, the behavior becomes nonwetting. 
The cost of depositing a
film becomes too great for wetting to occur when the surface is so rough.

Fig. 7 displays  results indicating  the sensitivity of the adsorption to
the presence of holes of various size on the surface. Since a small hole
(i.e. a single atom) postpones the quasi-transition jump, it is not
surprising that a big hole postpones it even more. In the case of a very
wide and deep hole, however, the behavior is quite different. The hole then
is so big as to provide a very attractive environment for Ne, inducing
adsorption even at quite low $P$ because of the favorable coordination within
the corners of the large hole. Ultimately, at sufficiently high $P$, the
adsorption becomes similar to that on a flat surface in all cases shown.

Finally, Fig. 8 indicates the adsorption's dependence on the size of
islands of adatoms. As discussed above, the single atom yields a steeply
rising isotherm at pressure above the flat surface's transition value. 
A 2x2 adatom cluster produces a more drastic effect: the isotherm is rather
smooth, eventually rising to agree with that of the flat surface.

\section{Summary and conclusions}

This paper has explored the effect of heterogeneity on the wetting
transition. To the best of our knowledge, ours is the first such study for a
system which is both realistic and relevant to current or forthcoming
experiments. The case of Ne is one which has exhibited wetting transitions
on alkali metal surfaces. Unfortunately, these transitions are not amenable
to  our simulation method because they occur so close to the critical
temperature. This has led us to explore the case of Mg, for which the
wetting transition is predicted to occur at $\approx 60 \%$ 
of the bulk Ne critical temperature $(44.4\,\kelvin)$.

 Our most intriguing result is that the adsorption isotherm's {\it shape} is
relatively insensitive to small heterogeneity; a $2 \%$ (of surface atoms)
imperfection frequency leads to an isotherm which appears to be
discontinuous. Quantitatively, however, the
shift is considerable; the quasi-transition occurs at a pressure nearly 5 \%
different from that of the flat surface. The same qualitative trend occurs
for more irregular surfaces, as seen in Fig. 6. In contrast, some forms of
heterogeneity, e.g. bumps, lead to the kind of smooth and continuous
behavior which might have been expected generically; see Fig. 8.  Such
behavior manifests no remnant of a transition.

Our results are based on very simple models of the potential, surface
geometry, and varieties of heterogeneity. Nevertheless, we suspect that a
few provisional conclusions are valid. A general (and unfortunate)
conclusion is that adsorption data are not easily ``inverted'' to learn about
the nature or degree of heterogeneity. Indeed, we found several instances
where heterogeneity yielded sharp, transition-like behavior which might be
misinterpreted as arising from a perfect, or near-perfect, surface.
Finally, we have confirmed the expectation that heterogeneity eliminates
the first order wetting transition, at least in the present case.

Future theoretical and simulation work is needed to extend our study. One
important question is whether the so-called  ``quasi-transition'' behavior is
actually singular, reflecting a higher order transition. Another question
pertains to the behavior of the isotherms very close to the wetting
transition temperature. These problems, as well as pursuit of the critical
regime, will take much more simulation time than we have employed here. It
is clear that such a study will be very useful, as will be experimental
study of the sensitivity of the isotherms to surface structure.

This research has been supported by the National Science Foundation and a
grant to Ing. S. Curtarolo from Fondazione Ing. Aldo Gini. 
We are grateful to J. R. Banavar, V. Bakaev, R. Blossey, M. H. W. Chan, 
W. F. Saam, A. M. Vidales, A. L. Stella and G. Stan 
for helpful discussions and comments.



\newpage


\begin{center} CAPTIONS \end{center}
 
1a. Potential energy  $U(x,y)$, defined in the text, as a function of lateral
position on the surface for the case of one Mg adatom at 
$(x,y)=(8.02,8.02 {\rm \AA})$ and one missing surface atom at $(20.05,20.05 {\rm \AA})$
in each simulation cell, which contains 49 surface atoms. Scale at
right is expressed in units of the well depth of the gas-Mg atom pair
potential, which has the value $15\,\kelvin$.

1b. Potential energy $V(x,y,z)$ as a function of normal distance z above
three surface atoms' sites: above the adatom (dashes), above the pit
(dash-dot), and above a surface atom in the unperturbed surface (full
curve). The curves are shifted so that their minima coincide. The values
of $z_{min}$ are $9.50, 3.6$ , and  $5.16\,\angstrom$, respectively.

2. Adsorption isotherms on a flat Mg surface (full curve), a surface with a
single adatom per unit cell (short dash), a surface with one missing adatom
(long dash), and a surface with an adatom and a missing surface atom
(medium dash). The area of the Mg unit cell is $16.08$ \AA$^2$.

3. 2d density of adsorbed Ne on the surface of figure 1, at $P=0.945\,\atm$,
$N=47.4$ particles (0.96 particles per Mg unit cell), just below the rapid rise 
of adsorption. Density scale at
right is expressed in units of inverse Angstroms squared.

4. Same as figure 3, except at $P=0.965\,\atm$ and $N=180$ (3.67 particles per unit cell).

5. Same as figure 3, except at $P=1\,\atm$ and $N=237$ (4.83 particles per unit cell).

6. Adsorption isotherms on a flat surface (full curve) and on rough surfaces
with $\sigma$ value as follows: one missing surface atom ($0.08$, small dash),
six randomly distributed missing atoms  ($0.2$, medium dash), 
10 random missing atoms ($0.3$, large dash),
and 11 random missing atoms plus 6 adatoms ($0.4$, dash-dot). The last of these
shows a rapid increase in coverage only at svp, meaning it is a nonwetting
surface.

7. Adsorption isotherms in the case of a flat surface (full circles), a
single missing surface atom (small dash), a cubic hole of depth $8.02\,\angstrom$ (medium
dash), and a  cubic hole of depth $16.04\,\angstrom$  (large dash).

8. Adsorption isotherms in the case of a flat surface (full curve), a single
adatom (dots),  and a 2x2 quartet of adatoms (dashes).

\begin{figure}[!h]
  \begin{center} 
   {\bf FIG. 1a}
    \epsfig{file=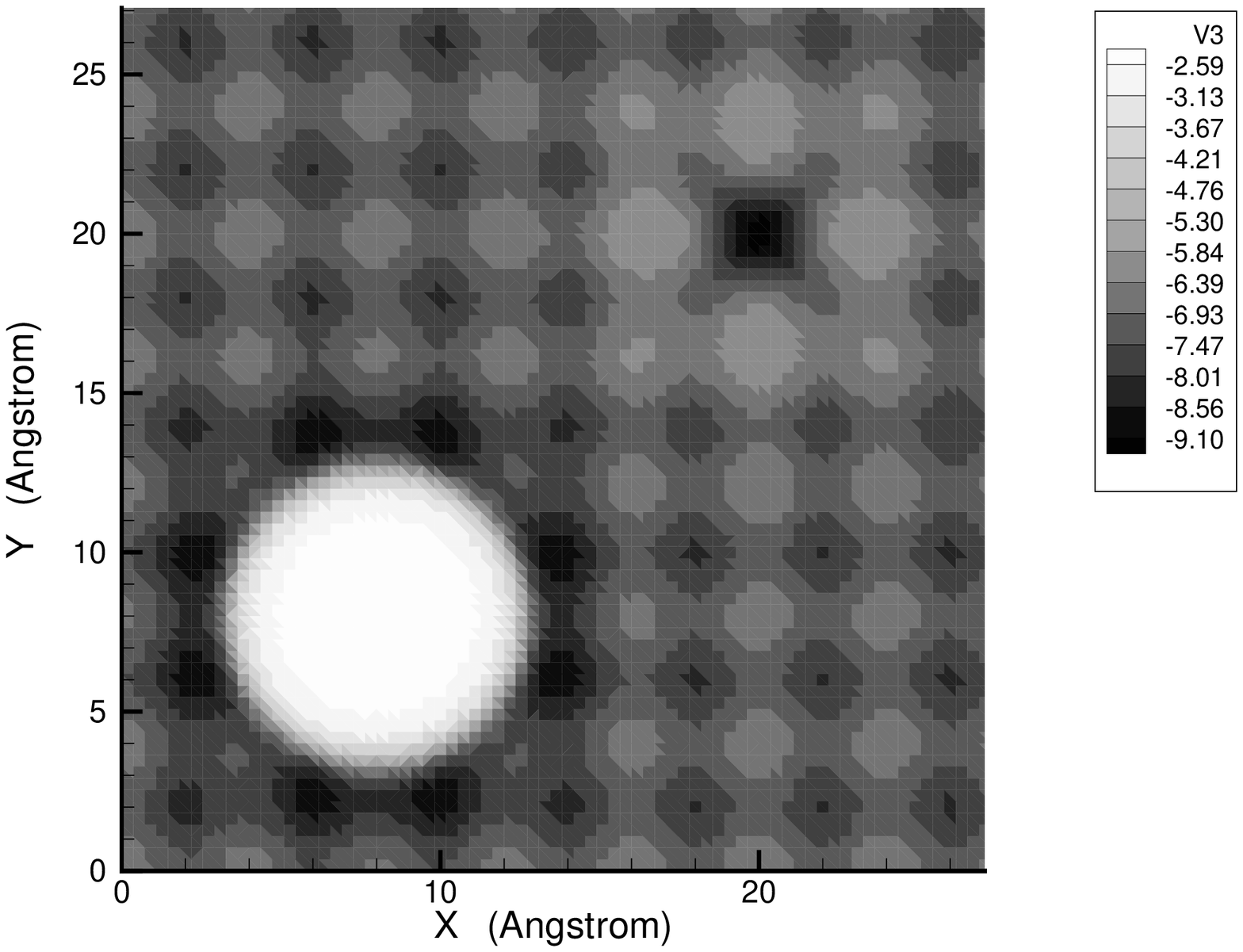}
  \end{center}
\end{figure}

\begin{figure}[!h]
  \begin{center} 
   {\bf FIG. 1b}
    \epsfig{file=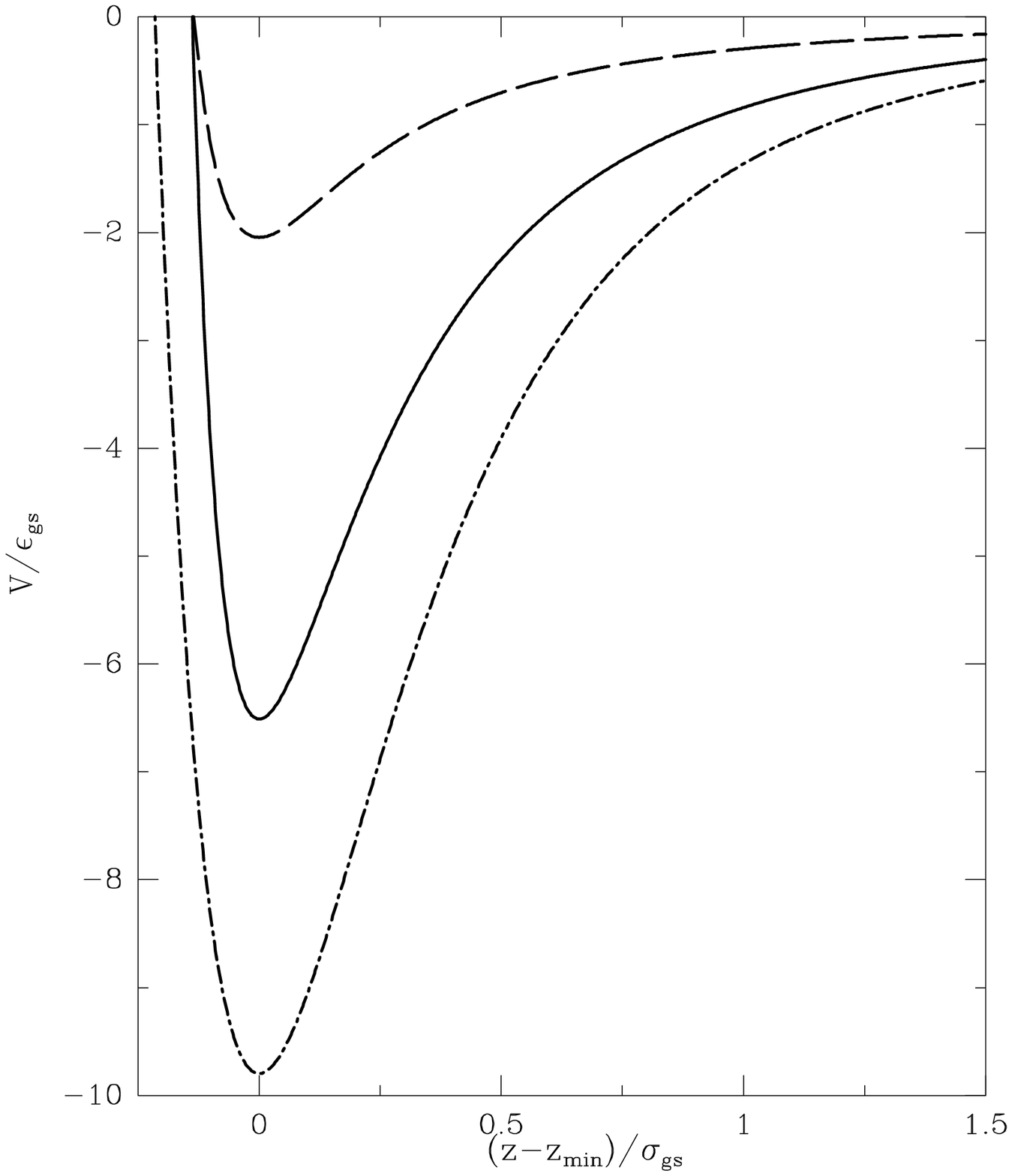,bbllx=0pt,bblly=0pt,bburx=30cm,bbury=30cm}
  \end{center}
\end{figure}

\begin{figure}[!h]
  \begin{center} 
   {\bf FIG. 2}
    \epsfig{file=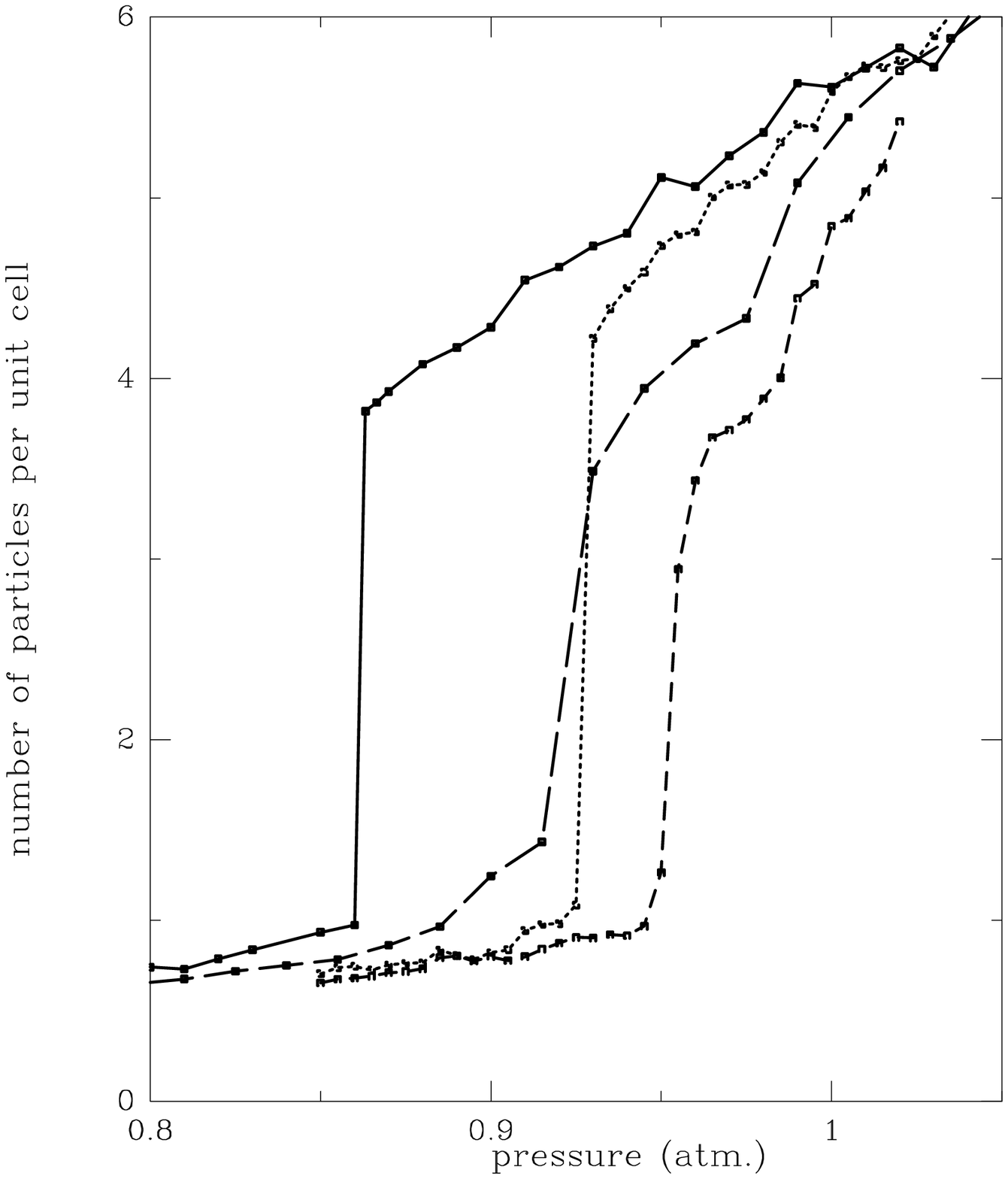,bbllx=0pt,bblly=0pt,bburx=30cm,bbury=30cm}
  \end{center}
\end{figure}

\begin{figure}[!h]
  \begin{center} 
   {\bf FIG. 3}
    \epsfig{file=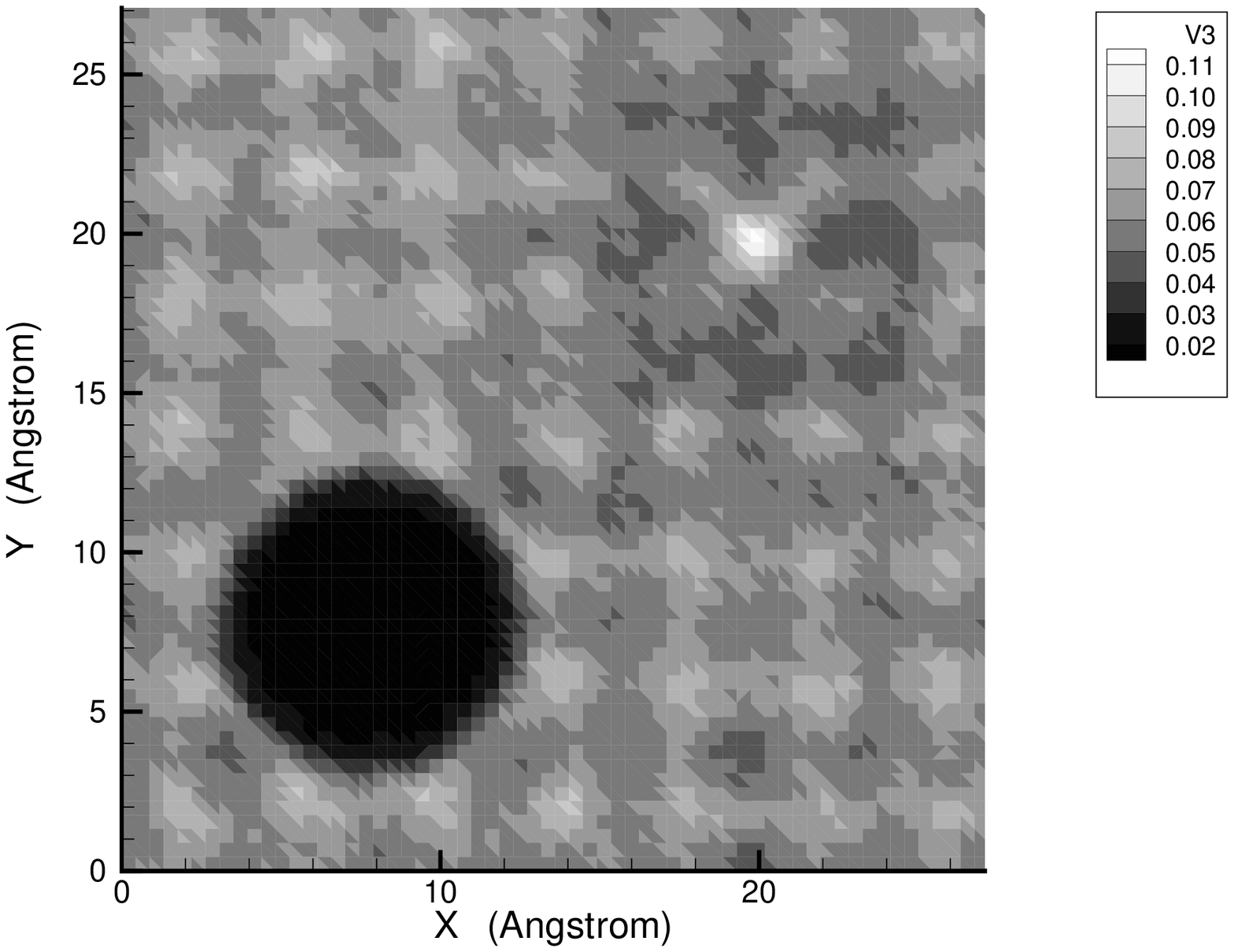}
  \end{center}
\end{figure}

\begin{figure}[!h]
  \begin{center} 
   {\bf FIG. 4}
    \epsfig{file=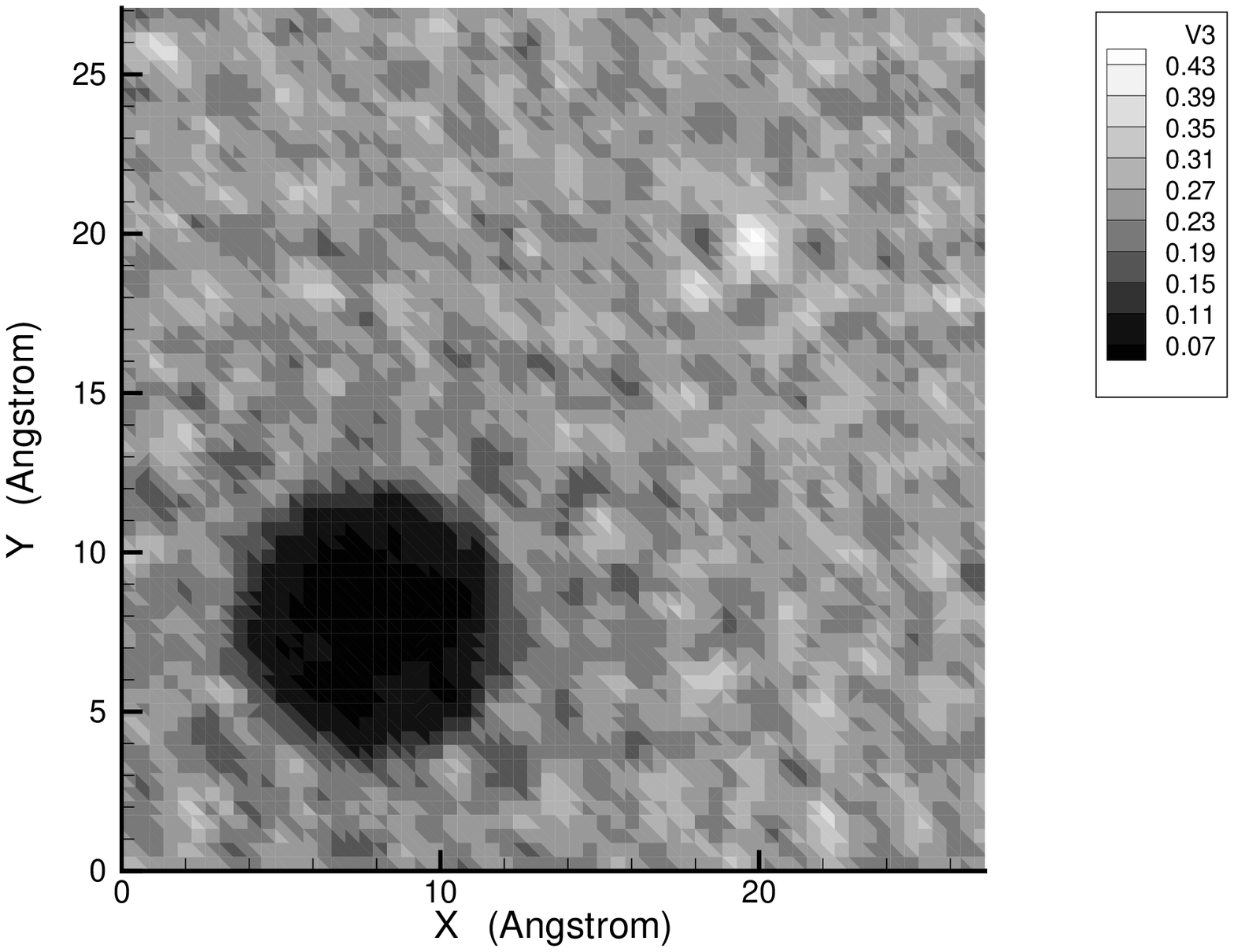}
  \end{center}
\end{figure}

\begin{figure}[!h]
  \begin{center} 
   {\bf FIG. 5}
    \epsfig{file=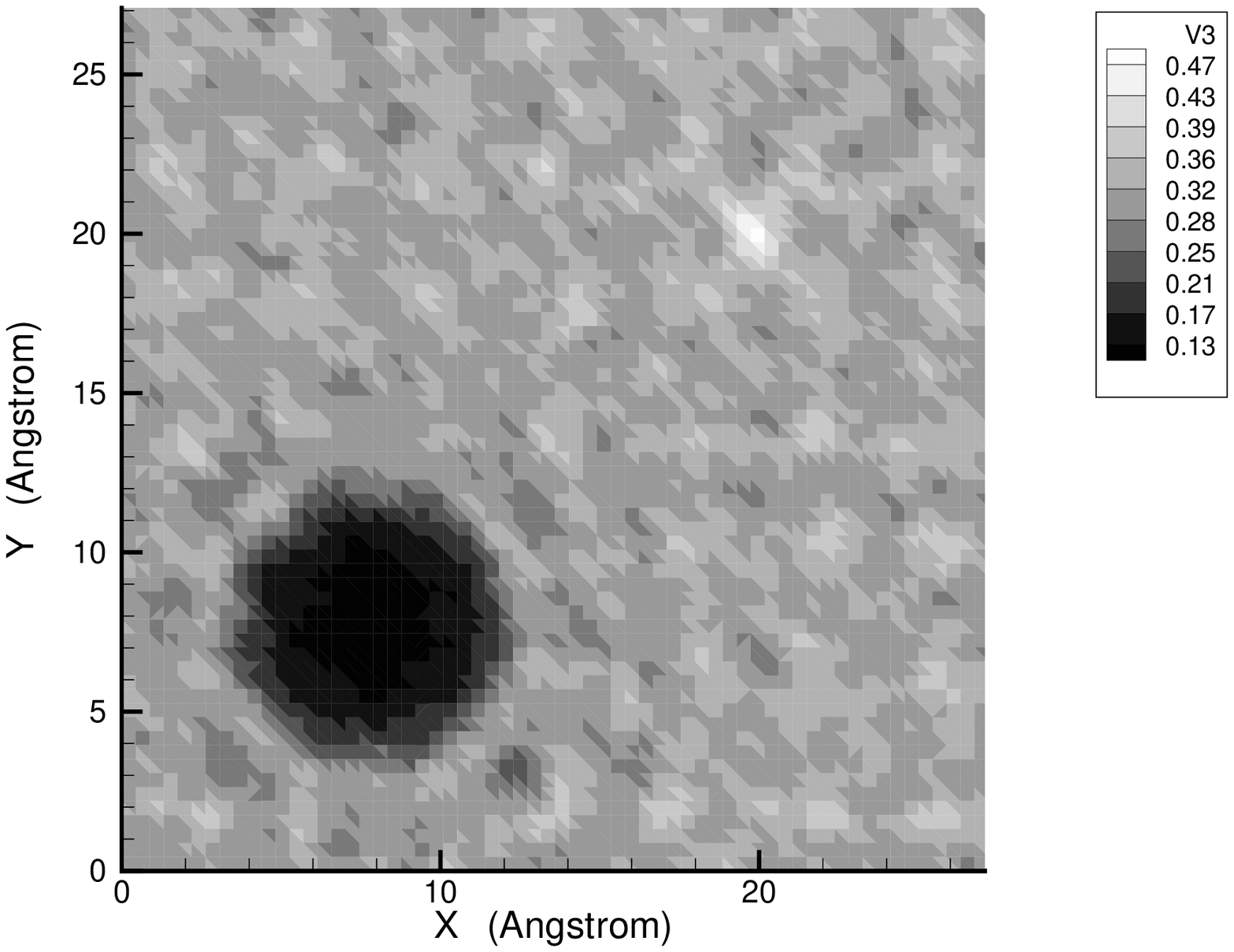}
  \end{center}
\end{figure}

\begin{figure}[!h]
  \begin{center} 
   {\bf FIG. 6}
    \epsfig{file=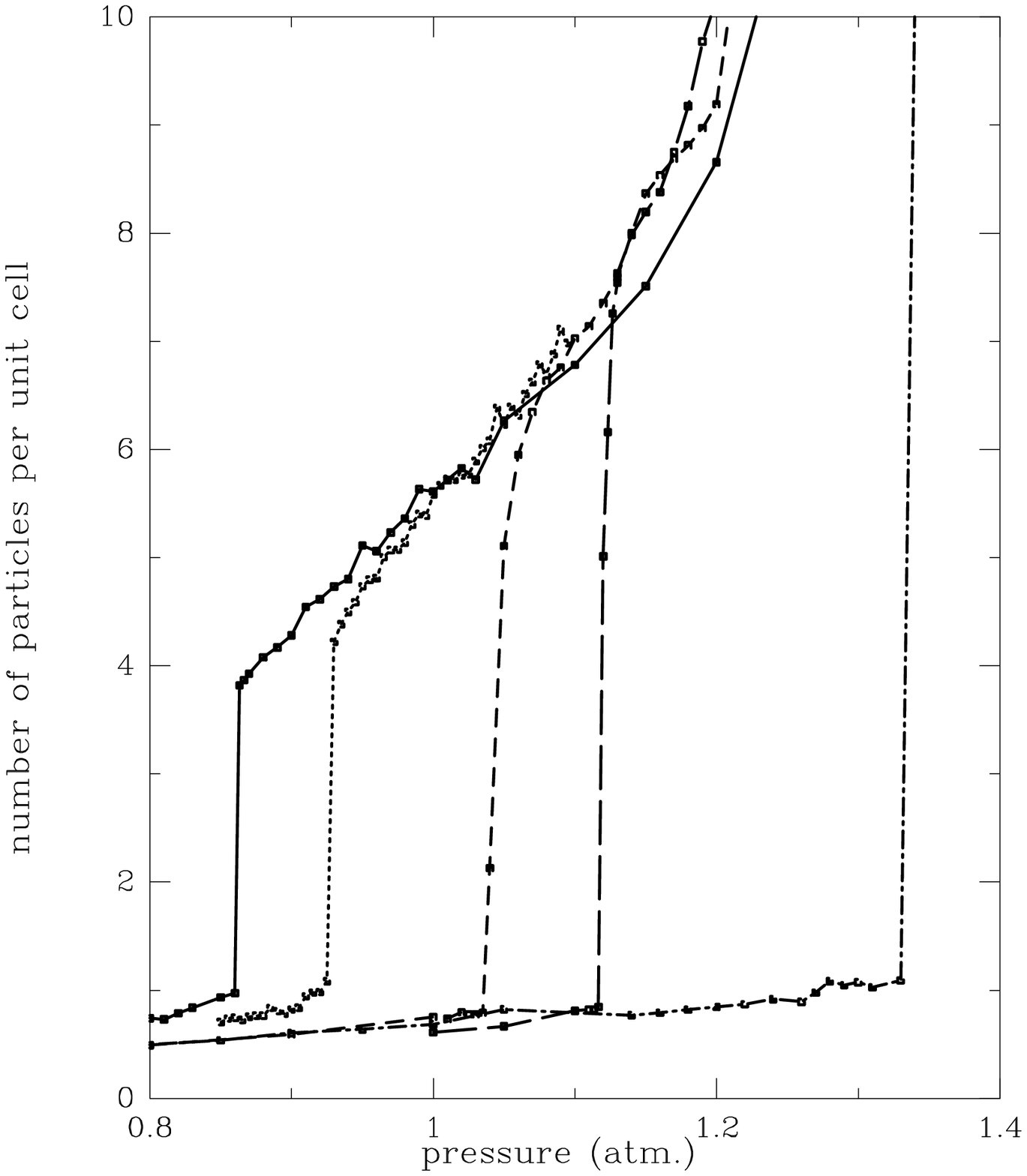,bbllx=0pt,bblly=0pt,bburx=30cm,bbury=30cm}
  \end{center}
\end{figure}

\begin{figure}[!h]
  \begin{center} 
   {\bf FIG. 7}
    \epsfig{file=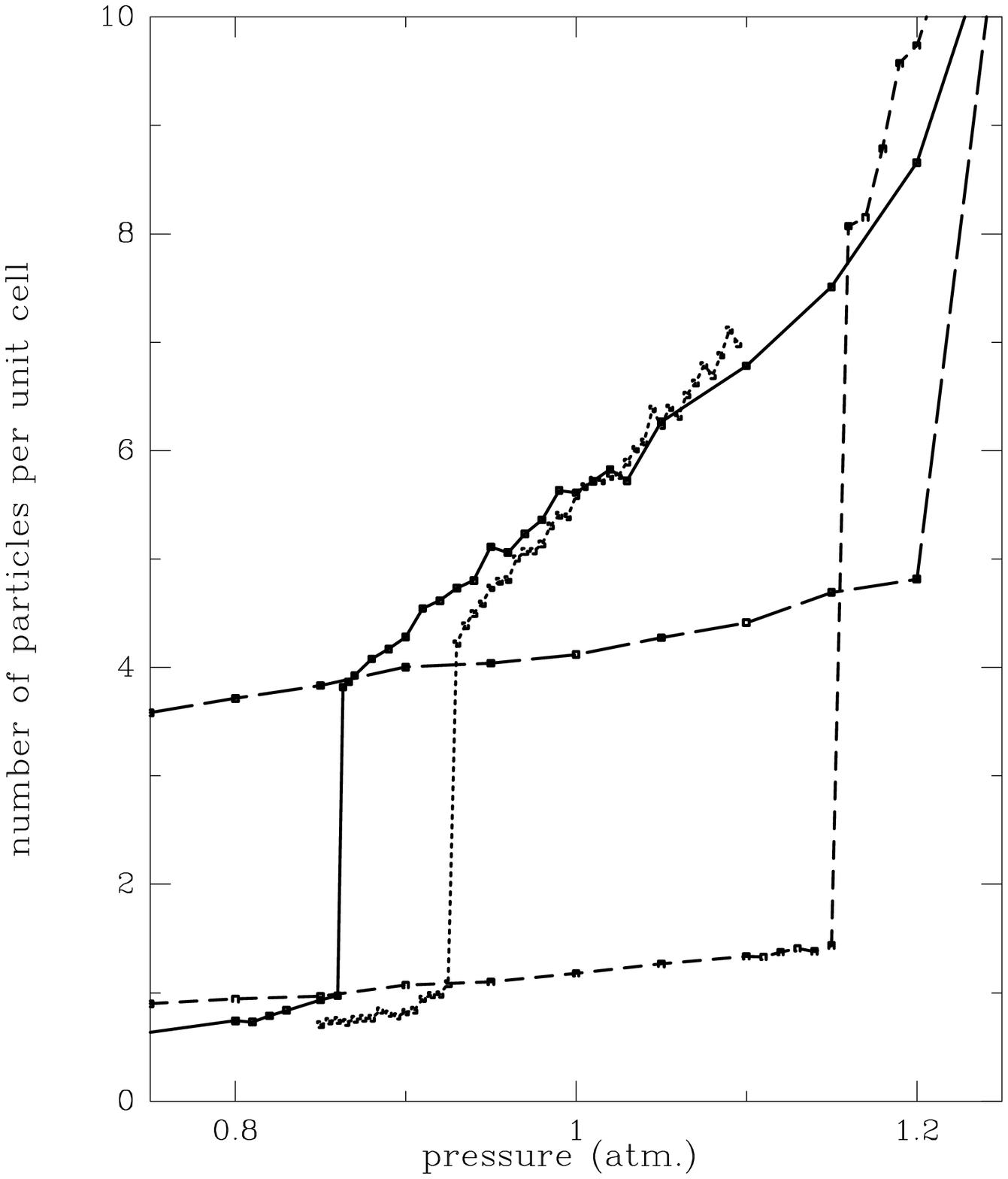,bbllx=0pt,bblly=0pt,bburx=30cm,bbury=30cm}
  \end{center}
\end{figure}

\begin{figure}[!h]
  \begin{center} 
   {\bf FIG. 8}
    \epsfig{file=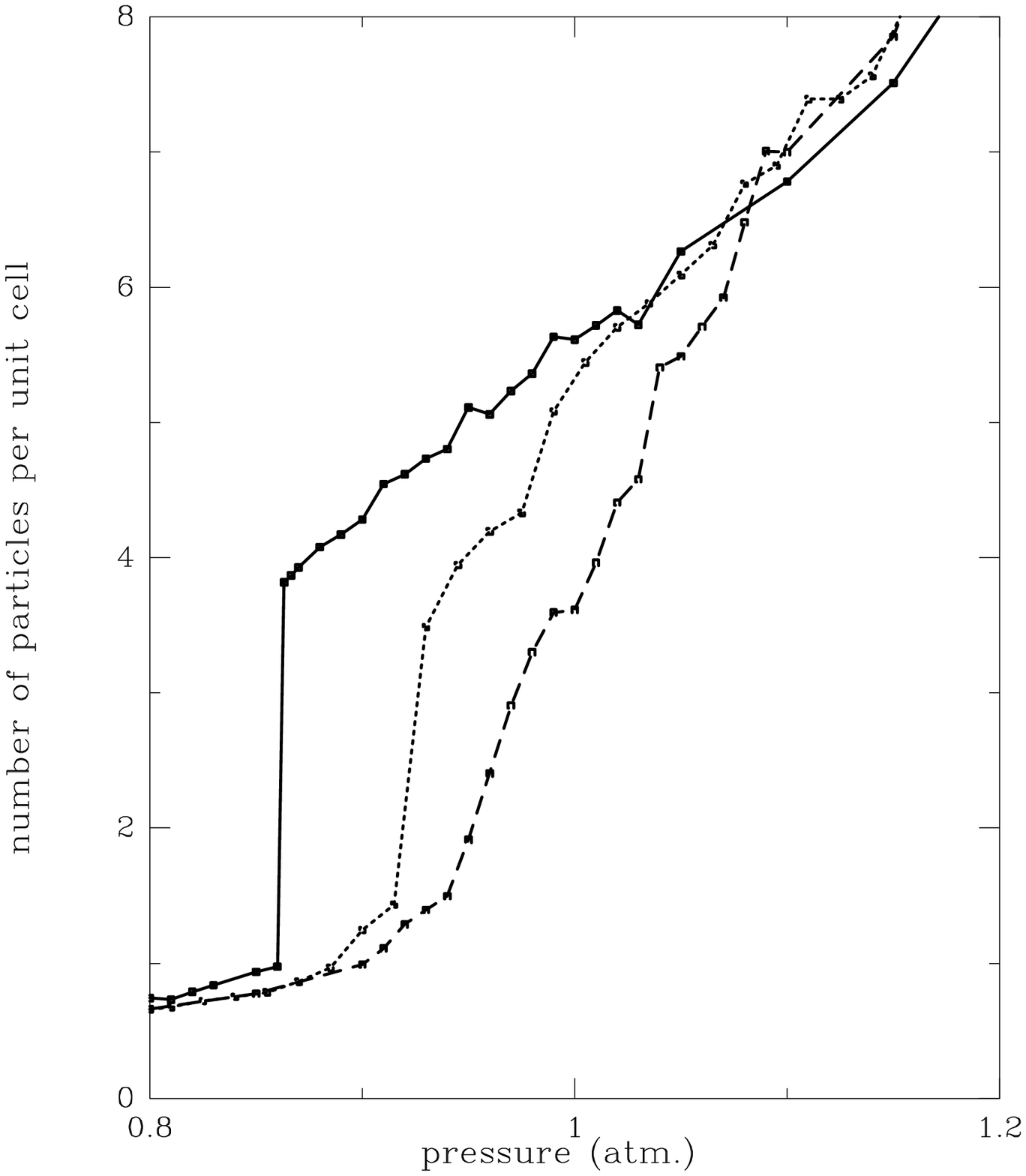,bbllx=0pt,bblly=0pt,bburx=30cm,bbury=30cm}
  \end{center}
\end{figure}


\begin{thebibliography}{99}
\bibitem{foot_1}
In the ``academic'' case of a surface consisting of weakly
interacting atoms or molecules, of course, one may add their individual
contributions for any arbitrary configuration.
\end{thebibliography}
\end{document}